\documentclass[aps, prd, twocolumn, lengthcheck, superscriptaddress, showpacs, letterpaper, nofootinbib]{revtex4-1}

\usepackage{epsfig}
\usepackage[usenames]{color}
\usepackage{graphicx}
\usepackage{amsmath}
\usepackage{epstopdf}
\usepackage{ulem}
\newcommand{\sbar}[1]{\ooalign{\hfil/\hfil\crcr$#1$}}

\def\bi{\bibitem}

\def\la{\langle}\def\ra{\rangle}
\def\be{\begin{eqnarray}}\def\ee{\end{eqnarray}}
\def\lsim{\mathrel{\rlap{\lower3pt\hbox{\hskip1pt$\sim$}}
     \raise1pt\hbox{$<$}}} 
\def\gsim{\mathrel{\rlap{\lower3pt\hbox{\hskip1pt$\sim$}}
     \raise1pt\hbox{$>$}}} 
\def\del{\partial}

\allowdisplaybreaks

\def\la{\langle}\def\ra{\rangle}
\def\be{\begin{eqnarray}}\def\ee{\end{eqnarray}}
\def\lsim{\mathrel{\rlap{\lower3pt\hbox{\hskip1pt$\sim$}}
     \raise1pt\hbox{$<$}}} 
\def\gsim{\mathrel{\rlap{\lower3pt\hbox{\hskip1pt$\sim$}}
     \raise1pt\hbox{$>$}}} 
\def\le{ \begin{array}{ll}}\def\re{\end{array}}

\def\lear{ \left( \begin{array}{cc}}\def\rear{\end{array} \right)}
\def\tr{\textnormal{tr}}

\def\le{ \left( \begin{array}{cc}}\def\re{\end{array} \right)}
\def\tr{\textnormal{tr}}

\def\bi{\bibitem}

\def\eft-hls{{\it EFT}$_{\rm bsHLS}$}
\def\skyrmion-hls{{\it Skyrmion}$_{\rm sHLS}$}

\begin{document}

\title{Nonquenching of  $g_A$ in nuclei, Landau-Migdal fixed-point theory, and emergence of scale symmetry in dense baryonic matter}

\author{Yan-Ling Li}
\affiliation{Center for Theoretical Physics and College of Physics, Jilin University, Changchun,
130012, China}

\author{Yong-Liang Ma}
\affiliation{Center for Theoretical Physics and College of Physics, Jilin University, Changchun,
130012, China}

\author{Mannque Rho}
\affiliation{Institut de Physique Th\'eorique,
	CEA Saclay, 91191 Gif-sur-Yvette c\'edex, France}
\date{\today}

\begin{abstract}
How the axial coupling constant $g_A$ in nuclear Gamow-Teller transitions described in shell model gets  ``quenched" to a universal constant close to 1 can be explained  by nuclear correlations in Fermi-liquid fixed-point theory using a scale-symmetric chiral Lagrangian supplemented with hidden local symmetric vector mesons. Contrary to what one might naively suspect --- and has been discussed in some circles --- there is no fundamental quenching at nuclear matter density due to QCD condensates.  When the density of many-body systems treated with the same Lagrangian increases beyond the density  $n=n_{1/2}\gsim 2n_0$ (where $n_0$ is the normal nuclear matter density) at which skyrmions representing baryons fractionize to half-skyrmions, with the $\rho$ meson driven toward the vector manifestation fixed point and a scalar meson $\sigma$ driven to the dilaton-limit fixed point with the nucleons parity-doubled, the dense matter supports the ``pseudoconformal" sound velocity for $n\gsim n_{1/2}$, while the trace of the energy momentum tensor remains nonvanishing. A plausible interpretation is that this signals the emergence of scale symmetry not explicitly present or hidden in QCD in the vacuum. The fundamental constant $g_A$, unaffected by QCD condensates  for $n< n_{1/2}$, does go to 1 as the dilaton-limit fixed point is approached before arriving at chiral restoration, but it is not directly related to the ``quenched $g_A$" in nuclei which can be explained as a Fermi-liquid fixed point quantity. The mechanism that produces a precocious pseudoconformal sound velocity is expected to  impact on the tidal deformability $\Lambda$ in gravitational waves from coalescing neutron stars.
\end{abstract}
\maketitle

\section{Introduction}

We explore in this paper in depth how the long-standing mystery of ``quenched" $g_A$ in nuclear Gamow-Teller transitions\footnote{We put the quotation mark to indicate that there is actually {\it no real quenching} as will be explained in this paper. In what follows, it should be understood as such although the quotation mark is not repeated.}  and the possible emergence of  symmetries hidden in QCD in dense baryonic matter relevant to massive compact stars can be intricately correlated. In this connection, the recent LIGO/Virgo detection of gravitational waves emitted from coalescing neutron stars GW170817~\cite{GW-Love} appears to be  poised to answer, among many of the most fundamental questions of Nature, the structure of highly dense matter in compact stars,  an outstanding problem both in astrophysics and nuclear physics. Accessible neither by perturbation nor by lattice --- the only known nonperturbative tool --- in QCD, there is at present no description that can be trusted of what  the structure is inside stabilized compact stars, presumably populated mainly by neutrons. In this paper, we attempt to describe how the property of dense compact-star matter can be correlated with what takes place in finite nuclei in terms of a topology change in the nucleon structure employing hadronic degrees of freedom only as density is increased beyond the normal nuclear matter density.

To address this problem, we take the effective field theory (EFT) framework developed in Ref.~\cite{bshls} to get the equation of state (EoS) for compact stars with an effective scale-chiral Lagrangian referred to as $bs$HLS, defined below in Eq.~(\ref{LOSS}). In this Lagrangian, the vector mesons $\rho$ and $\omega$ are introduced as hidden local gauge fields and a scalar dilaton field $\sigma$ figures as a pseudo-Nambu-Goldstone field in baryonic nonlinear $\sigma$ model. The effective Lagrangian will have its ``bare" parameters endowed with density dependence inherited from QCD so that Wilsonian-type renormalization-group treatment can be made in the ``sliding vacuum" that follows the density characterizing the many-baryon system (see, e.g., Refs.~\cite{HY:PR,Brown:2003ee} for detailed discussions.\footnote{It is perhaps appropriate to clarify here the Lorentz structure of density dependence that is involved in the ``bare" parameters resulting from the matching. Here coming from the QCD proper, the density dependence lies in the QCD condensates that are --- by construction --- Lorentz-invariant. As will be discussed in Section \ref{induced}, the spontaneous breaking of Lorentz symmetry in matter induces additional Lorentz-noninvariant density dependence, dubbed DD$_{\rm induced}$. Thus, ``effective parameters" that figure in the calculations will contain both the IDD and DD$_{\rm induced}$. } This framework has been found to describe fairly well the bulk property of normal nuclear matter as well as that of neutron stars, including  the observed massive $\approx 2$-solar mass object. That the approach of Ref.~\cite{bshls} does this feat is not a big deal by itself given that there are a large number of phenomenological models in the literature with numerous adjustable parameters that do similar or even better job. What is, however, distinguishable of this approach is that it relies on a topology change associated with the baryon structure, presumably encoding a smooth cross-over transition from hadrons to  QCD variables, and makes, with a fewer number of parameters, certain predictions that are not found in conventional approaches.

Among various predictions, the most striking one is that at the density denoted as $n_{1/2}$ where the topology change takes place from skyrmions to half-skyrmions in the solitonic description of baryonic matter~\cite{park-vento}, the symmetry energy $E_{\rm sym}$ in the EoS  develops a cusp~\cite{LPR}.
Notably, the EoS,  soft for  $n < n_{1/2}$, stiffens appreciably above $n_{1/2}$ and accounts  for  the massive compact stars observed in nature. Furthermore, the sound velocity of star $v_s$ approaches what is usually referred to as  ``conformal velocity" $v_s^2=1/3$ (in the unit of $c=1$)  associated with conformal symmetry, indicating a possible emergence in dense matter of conformal invariance  broken in the vacuum by the trace anomaly. Since the trace of the energy-momentum tensor in the medium does not vanish,  it is more appropriate to call it  ``pseudoconformal" sound velocity.  It also offers the mechanism to zero-in on  the tidal deformability $\Lambda$ in gravitational waves from coalescing stars as recently observed~\cite{GW-Love}.

A surprising spin-off of the effective field theory approach to dense matter relevant to compact stars mentioned above is that it reveals insight into what happens even at the normal nuclear matter density $n_0$, namely, that the same underlying Lagrangian with the scale-chiral symmetry structure, when suitably treated in the mean field with Landau Fermi-liquid fixed point theory,   can explain in an extremely simple way why the effective coupling constant $g_A^{\rm eff}$ in nuclear Gamow-Teller transitions at  zero-momentum transfer described in {\it simple shell model} is ``quenched" to $g_A^{\rm eff}\approx 1$.  The result, which corresponds to a Fermi-liquid fixed point quantity, encapsulates  nuclear core-polarizations~\cite{corepolarization} and coupling to $\Delta$-hole states~\cite{delta-hole,Delta-GTR,bohr-mottelson,auerbach}, both strongly mediated by the tensor forces and other effects suppressed by the chiral counting such as two-body meson-exchange currents.

\section{Topology change}\label{topology}

\begin{figure}[h]
\begin{center}
\includegraphics[width=8cm]{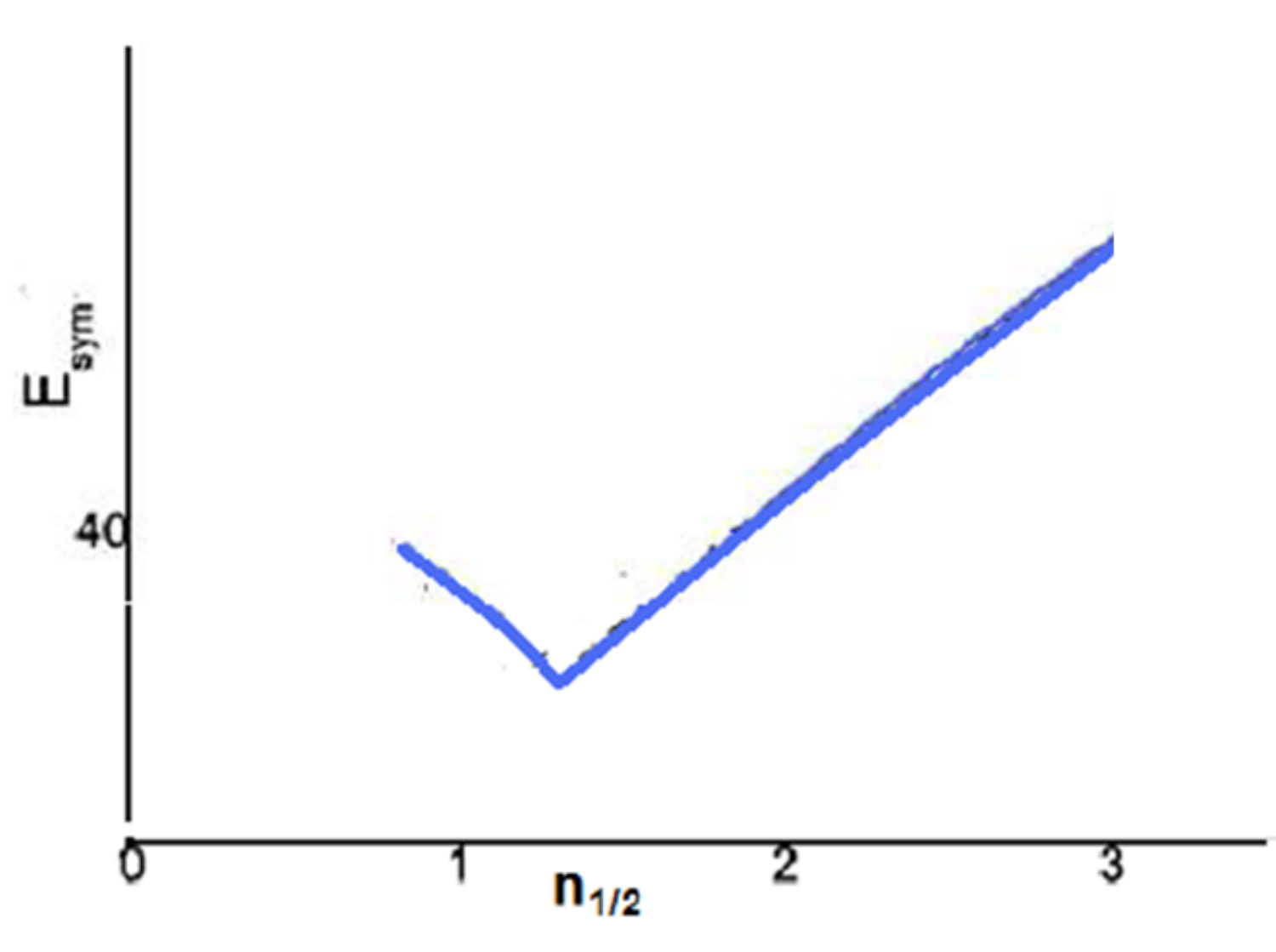}
\caption{ Schematic cusp structure for $E_{\rm sym}$ reproduced from Ref.~\cite{LPR}. The energy and density scales are arbitrary. What is relevant is the generic cusp structure which depends only on topology, hence only on the pion field.}\label{cusp}
\end{center}
\end{figure}

Arguably the most important  quantity in the EoS for compact stars is the symmetry energy $E_{\rm sym}$ defined as the difference between the pure neutron matter and symmetric nuclear matter in the energy per particle of the system,
\be
E(n,\alpha)=E_0(n,0)+E_{\rm sym}\alpha^2 +\cdots
,
\ee
where $\alpha=(N-Z)/(N+Z)$.  A particularly interesting information on $E_{\rm sym}$ comes from baryonic matter described in terms of skyrmions as baryons, valid in the large $N_c$ limit.
For this an efficient tool is to put skyrmions on crystal lattice~\cite{park-vento}, the favored lattice configuration being the face-centered cubic crystal. If one quantizes the skyrmion system by rotating in modular space, one finds the rotation energy  to be given by~\cite{LPR}
\be
\Delta E=(8{\cal I})^{-1}\alpha^2
,
\ee
which gives
\be
E_{\rm sym}=(8{\cal I})^{-1}
,
\ee
where ${\cal I}$ is the isospin moment of inertia.  Given the skyrmion configuration in medium, this expression can be easily computed. It is topological, leading order in $1/N_c$ and  is considered to be robust. Other degrees of freedom than the pion --- which carries topology --- do not influence the qualitative feature, namely, the appearance of the cusp structure
at a density $n_{1/2}$ at which there is a changeover from skyrmions to half-skyrmions. This is represented pictorially in Fig.~\ref{cusp}.

The cusp can be readily understood by the fact that the skyrmion with unit baryon number in the phase $n<n_{1/2}$ fractionizes at  $n=n_{1/2}$ into two half-skyrmions, each of which carrying half baryon charge. This changeover is favored by the symmetries involved~\cite{goldhaber-manton,park-vento}. The half-skyrmions, however, are not propagating degrees of freedom since they are confined by monopoles. But they play a crucial role for the structure of the state in the sector $n\geq n_{1/2}$.  With the half-skyrmion structure, the chiral condensate characterized by the bilinear quark condensate $\la\bar{q}q\ra$  vanishes when averaged over the space, $\Sigma\equiv \overline{\la\bar{q}q\ra}=0$, but it is locally nonzero, hence supporting  chiral density wave in the system. We will see below this structure plays a very important role. Now as $\Sigma$ goes to zero approaching $n_{1/2}$ from below,  terms multiplied by $\Sigma$ go to zero but what remains nonzero increases and that produces the cusp. One may think this as an artifact of the lattice structure. But it is not as we will argue below.

\section{Vector Manifestation}

\begin{figure}[h]
\begin{center}
\includegraphics[width=8.5cm]{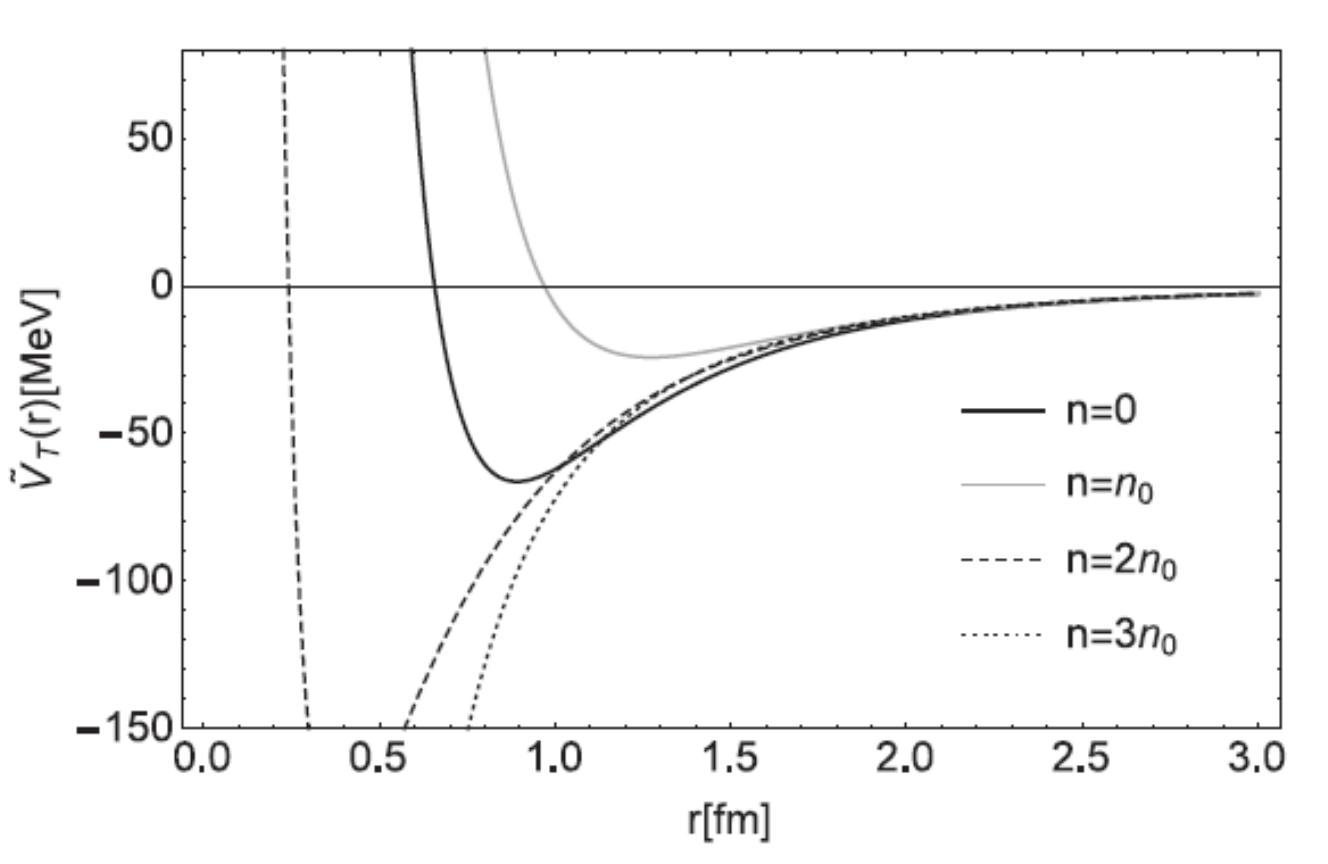}
\caption{ $\tilde{V}_T (r) =V_T (r) (\tau_1\tau2 S_{12})^{-1}$ at various densities $n$~\cite{bshls}. Note that the $\rho$-tensor force is suppressed almost completely at $n\gsim 2n_0$.}\label{tensor}
\end{center}
\end{figure}

In fact, it is easy to reproduce this cusp structure using a standard nuclear physics argument with the baryonic hidden local symmetry (HLS) Lagrangian.  For this, we transcribe the topology change in the skyrmion structure into the ``bare" parameters of the Lagrangian $bs$HLS that are sliding with density. Here what figures most importantly is the property of the $\rho$ meson at high density. In hidden local symmetry (HLS), Wilsonian renormalization-group analysis predicts that as the quark condensate is driven to zero, say, by density, the vector coupling $g_\rho\to 0$, hence $m_\rho$ which is given by the KSRF relation~\footnote{The KSRF relation is a low-energy theorem which holds to all orders of loop corrections and becomes exact as $m_\rho$ tends to zero.}  should go to zero independently of the pion decay constant $f_\pi$~\cite{HY:PR}. This is known as the ``vector manifestation (VM)" phenomenon. In $bs$HLS, the nuclear tensor force is given at the leading chiral order  by  the sum of one-$\pi$ and one-$\rho$ exchanges with the parameters with {intrinsic density dependence (IDD)} sliding with density {as specified below}. Due to the cancellation between the two tensor forces, the (net) tensor force in the range of nuclear forces relevant for the nuclear interaction, $r\gsim 1$ fm, tends to drop in strength when the density is approach to $n_{1/2} = 2n_0$ from below, and nearly vanishes at a density near 2 $n_0$~\cite{bshls}.  This feature is indicated in Fig.~\ref{tensor}.

This decrease of the tensor force has been observed in certain spin-isospin-dependent processes in heavy nuclei and has given a beautiful explanation for the long life-time of  the C14 beta transition, zero-ing on the decreasing tensor force near $n_0$~\cite{holt}. The dropping of the (net) tensor force stops abruptly at $n_{1/2}$ because of the onset of the vector manifestation which makes $g_\rho$ drop rapidly, strongly suppressing the $\rho$ tensor force. The pion tensor force,  more or less independent of density,  takes over at this density and increases as density goes up,  so that  the net tensor force turns up and moves upward. What happens then can be easily seen by using the closure approximation that is valid near $n_0$,
{
\be
 E_{\rm sym}\approx \la |V^T|^2\ra/\bar{E}
 \ee
 }
where $\bar{E}\approx 200$ MeV, the mean energy of the states to which the tensor force connects strongly from the ground state. The decrease of the tensor force  going toward $n_{1/2}$ and the rapid increase due to the $\pi$ tensor after $n_{1/2}$ gives the cusp. This feature reproduces precisely the cusp of Fig.~\ref{cusp}.

\begin{figure}[ht]
\begin{center}
\includegraphics[width=8cm]{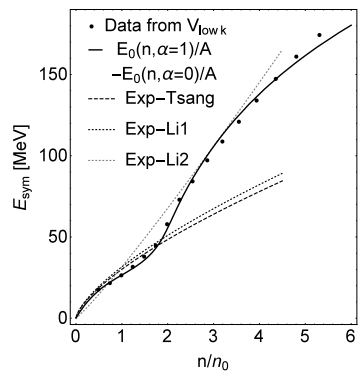}
\caption{The symmetry energy $E_{\rm sym}$ in $V_{\rm lowk}$ RG (dots) reproduced from Ref.~\cite{bshls} (for the notations, see the explanations therein).
 }\label{Esym}
 \end{center}
\end{figure}

The closure approximation is at best a rough estimate. With the given Lagrangian with the sliding vacuum structure incorporated into $bs$HLS, one can do a sort of Wilsonian RG calculation. This is described in Ref.~\cite{bshls} using what is called $V_{\rm lowk}$. High-order nuclear correlations included in the RG flow via $V_{\rm lowk}$, which represent higher order $1/N_c$ corrections to the skyrmion cusp and $1/\bar{N}$  Fermi-liquid corrections to the closure-sum approximation cusp where $\bar{N}=k_F/(\Lambda_F -k_F)$ and $\Lambda_F$ is the $V_{\rm lowk}$RG cutoff,  smoothen the cusp  and give the soft-to-hard transition in the EoS as given in Fig.~\ref{Esym}. That changeover  can account for $\approx 2$-solar mass stars. Although smoothed,  the cusp structure remains as we will see how it impacts on the tidal deformability described below.

%
\section{Scale-chiral symmetry}
Thus far we have not specified how the scalar meson of mass $\approx 600$ MeV which plays an extremely important role in nuclear physics enters in nuclear dynamics. As distinguished from the way the scalar degree of freedom figures in the standard relativistic mean field approaches, we introduce the scalar meson as a dilaton, a pseudo-Nambu-Goldstone boson of both explicitly and spontaneously broken scale symmetry. For this we follow the model proposed by Crewther and Tunstall (CT)~\cite{CT,CCT}. We wish to incorporate in addition the hidden local symmetric mesons $\rho$ and $\omega$ into the scale-symmetric chiral perturbation  Lagrangian $\chi$PT$_\sigma$ constructed by CT. Such a Lagrangian that hereon we shall refer to as $s$HLS  was written down to next-to-leading order {(NLO)} in scale-chiral counting in Ref.~\cite{LMR}.

To define the notations and also to be self-contained, we briefly review the procedure, the detail of which is found in Ref.~\cite{LMR} and references given therein.  First we start with the HLS Lagrangian ${\cal L}_{\rm HLS}$  written to the leading chiral order $O(p^2)$. To that order, the HLS Lagrangian --- apart from the chiral-symmetry breaking quark mass term --- transforms under scale transformation $x\to \lambda^{-1}x$ as
\be
{\cal L}_{\rm HLS;LO} (x)\to \lambda^2 {\cal L}_{\rm HLS;LO} (\lambda^{-1} x).
\ee
Next the action $\int d^4x {\cal L}_{\rm HLS}$ is made scale invariant by multiplying the Lagrangian density by the conformal compensator field $\chi$ that scales as mass dimension 1 and scale dimension 1 so that the resulting action is scale-invariant,
\be
{\cal L}_{s\rm HLS;LO}\equiv \left(\frac{\chi}{f_\sigma}\right)^2{\cal L}_{\rm HLS;LO}
,
\ee
where $\chi=f_\sigma e^{\sigma/f_\sigma}$ with $\sigma$ being identified as dilaton transforming nonlinearly under scale transformation and $f_\sigma$ the $\sigma$ decay constant. In the CT formalism, the effective Lagrangian in the leading order contains the anomalous dimension of $\tr (G_{\mu\nu}^2)$, $\beta^\prime$, the slope of the $\beta$ function for $\alpha_s$ given at the putative IR fixed point $\alpha_{\rm IR}$ as
\be
{\cal L}_{\rm HLS;LO}^{\beta^\prime, c_{\rm hls}}=\left(c_{\rm hls}+(1-c_{\rm hls})\left(\frac{\chi}{f_\sigma}\right)^{\beta^\prime}\right){\cal L}_{\rm HLS;LO}
\label{CTform}
\ee
in each term in ${\cal L}_{\rm HLS;LO}$,  where $c_{\rm hls}$ is an arbitrary constant. Note that
 $\beta^\prime$ is $O(p^0)$ in the scale-chiral counting. Also the kinetic energy term for the dilaton is made up of two terms,
\be
{\cal L}_{\rm dilaton}^{\beta^\prime,c_\sigma}=\Big(c_\sigma +(1-c_\sigma)\big(\frac{\chi}{f_\sigma}\big)^{\beta^\prime}\Big)\frac 12 \del_\mu\chi\del^\mu\chi
,
\ee
where $c_\sigma$ is yet another arbitrary constant. The leading scale-chiral order Lagrangian then is
\be
{\cal L}_{\rm LO}={\cal L}_{\rm HLS;LO}^{\beta^\prime, c_{\rm hls}}+{\cal L}_{\rm dilaton}^{\beta^\prime,c_\sigma} + V_{m} +V_\chi
,
\ee
where $V_m$ is the chiral symmetry breaking quark mass term and $V_\chi$ the dilaton potential that triggers both explicit and spontaneous breaking of scale symmetry. One can go on to higher orders in the scale-chiral counting. In Ref.~\cite{LMR}, the scale-symmetric HLS Lagrangian has been written down up to NLO  including baryons.

Already at the leading order, the proliferation of presently unknown parameters, such as $\beta^\prime$, $c_i$'s, $\gamma_m$ --- the anomalous dimension of the quark mass term --- etc., appear to make the resulting Lagrangian uncontrollable, not to mention  applications to nuclear matter and dense matter. At the NLO, this gets humongous with some 50 unknown parameters~\cite{LMR}. It turns out, however, that it is not hopeless. There turns out be a remarkable simplification that one can make without losing predictive power in applying to nuclear systems. It is what we refer to as ``leading-order-scale symmetry" (LOSS) approximation. It amounts to taking in Eq.~(\ref{CTform})
\be
c_k\approx 1\ {\rm  for\ all}\  k.
\ee
This allows $\beta^\prime$ to be nonzero but removes  the $\beta^\prime$ dependence in dilaton-matter coupling in the leading-order Lagrangian,  leaving its presence entirely in the potential $V_\chi$ --- and also in $V_m$ if the quark mass is taken into account. This approximation is consistent with the argument by Yamawaki that scale symmetry is in fact hidden in the standard model~\cite{yamawaki}. Yamawaki develops the argument using  the linear $\sigma$ model to which the standard Higgs model is equivalent. It is shown that by dialing a constant, the linear $\sigma$ model can be brought in the strong coupling limit to nonlinear sigma model consistent with low-energy chiral Lagrangian, while it is brought,  in the weak coupling limit, to a scale-invariant Lagrangian with scale-symmetry breaking lodged in a potential. The structure of the LOSS Lagrangian is similar to this limit, modulo the dilaton potential which must depend on the basic structure of scale symmetry breaking, which is unknown in QCD with $N_f = 3$. A similar structure of a scale-chiral Lagrangian was obtained in the leading order by Golterman and Shamir (GS)~\cite{GS} who invoked the Veneziano limit wherein both $N_c$ and $N_f$ are taken to infinity with $N_f/N_c$ kept finite to construct scale-chiral effective field theory in the vicinity of the ``sill" to a conformal window. Up to NLO, the two formulations, CT and GS, approaching  a putative IR fixed point from below, although their IR fixed-point structures are basically different~\cite{CCT}, have the same scale-chiral perturbative structure.  In what follows the point of view we adopt, our basic assumption, is that regardless  of whether an IR fixed point is present at $N_f= 3$ in QCD in the vacuum, scale symmetry could ``emerge" in dense baryonic matter as density is dialed up and it is the weak coupling limit \`a la Yamawaki  that seems to be sampled by dense matter.

The Lagrangian in the LOSS approximation we use in what follows is of the form
 \be
 {\cal L}_1= {\cal L}^{M}_{\chi PT_\sigma} (\pi,\chi,V_\mu)+ {\cal L}^{B}_{\chi PT_\sigma} (\psi,\pi,\chi,V_\mu) +V(\chi).\nonumber\\
 \label{LOSS}
 \ee
 The first is the $s$HLS
 \be
 {\cal L}^{M}_{\chi PT_\sigma} (\pi,\chi,V_\mu) & = & \frac{1}{2}\partial_\mu \chi \partial^\mu \chi+ f_\pi^2\left(\frac{\chi}{f_{\sigma}}\right)^2{\rm
Tr}[\hat{a}_{\perp\mu}\hat{a}_{\perp}^{\mu}] \nonumber\\
& &{} + a f_\pi^2 \left(\frac{\chi}{f_{\sigma}}\right)^2 {\rm Tr} [\hat{a}_{\parallel\mu}\hat{a}_{\parallel}^{\mu}]
\nonumber\\
& &{} -\frac{1}{2g^2}{\rm Tr}[V_{\mu\nu}V^{\mu\nu}] , \label{shls}
 \ee
 and the second is the baryonic coupling to $s$HLS fields
 \be
 {\cal L}^{B}_{\chi PT_\sigma} (\psi,\pi,\chi,V_\mu) & = & {\rm Tr} (\bar{B}i\gamma_\mu D^\mu B)-\frac{\chi}{f_{\sigma}} m_N {\rm Tr}(\bar{B}B)\nonumber\\
 & &{} +\cdots \label{bshls}
 \ee
and $V(\chi)$ is the dilaton potential, the explicit form of which is not needed for our purpose.
\section{Energy density functional}
\subsection{Sliding vacua}

Dense matter described in terms of skyrmions, while difficult to quantize, provides highly nontrivial information. We exploit it to gain what can be topologically robust inputs as a function of density. In the large $N_c$ limit and at high density, this should be reliable. This is done by putting skyrmions on crystal lattice and varying the crystal size~\cite{park-vento}. The background given by the skyrmion matter then provides an ``intrinsic density dependence (IDD)" --- to be defined precisely below --- considered to be equivalent to that inherited from QCD by the matching of the correlators as described more precisely below to the fluctuating  degrees of freedom involved in the dense system. Those informations obtained from the skyrmion crystal are then transcribed into $bs$HLS derived by incorporating baryons into $s$HLS in consistency with scale symmetry and hidden local symmetry~\cite{bshls}.

Our next step is  to embed $bs$HLS given in the LOSS approximation, Eq.~(\ref{LOSS}), in baryonic medium characterized by density $n$ and treat dense matter in a way familiar in nuclear physics as the ``energy density functional" theory, analogous to Kohn-Sham density functional theory  in atomic and chemical physics~\cite{kohn-sham}. In the mean-field approximation with Eq.~(\ref{bshls}), our approach will be  of the type known as ``relativistic mean field theory" \`a la Walecka. In terms of Wilsonian renormalization group approach, the mean-field approximation would correspond to  Landau Fermi-liquid fixed point theory. Going beyond the mean field including higher correlations is achieved by the $V_{\rm lowk}$RG that involves double decimations~\cite{bshls}. Indeed the double-decimation $V_{\rm lowk}$RG approach --- that we will adopt below --- amounts to  doing Landau (or more properly Landau-Migdal) Fermi-liquid theory as formulated in Ref.~\cite{SB-fermiliquid}.

For energy density functional  theory, as in Kohn-Sham's density functional theory, the density plays a key role. To arrive at what could correspond in nuclear dynamics to the Hohenberg-Kohn potential of density functional theory~\cite{hohenberg-kohn}, we match the correlators of the EFT Lagrangian Eq.~(\ref{LOSS}) at tree order to those of QCD operator-product expansion at a suitable matching scale as described in Ref.~\cite{bshls}. This allows the effective field theory to inherit certain important intrinsic properties of QCD coming from the scale above the matching scale.  This is highly intricate and difficult matter so we cannot claim what we are doing is accurate. What it does is that it renders nuclear many-body systems treatable in ``sliding vacua," making  the ``bare" parameters of Eq.~(\ref{LOSS}) endowed with the density-dependent condensates $\la\chi\ra^\ast$, $\la\bar{q}q\ra^\ast$, etc., where $\ast$ stands for density dependence. This density dependence inherited from QCD is denoted as ``intrinsic density dependence (IDD)" to be distinguished from density dependence coming from mundane many-body nuclear interactions.

\subsection{Skyrmion-half-skyrmion topology change}
\label{parity-doubling}

Consider dense matter in terms of skyrmions put on crystal lattice.  In addition to the cusp structure at the density $n_{1/2}$ in the symmetry energy discussed above,  the crossover to the half-skyrmion ``phase"\footnote{As noted, there is no order parameter change in going from skyrmions to half-skyrmions, hence  there is no bona-fide phase transition belonging to the Landau-Ginzburg-Wilsonian paradigm.  We shall nevertheless use this term with this understanding.} makes the effective mass of the baryon become a density-independent constant which can be identified with the chiral invariant mass $m_0$ figuring in the parity-doubling model~\cite{detar-kunihiro}. Since parity doubling is not in QCD, this is a symmetry ``induced" in medium and not intrinsic. Since the skyrmion crystal  is a reliable description at high density, if the topology change density $n_{1/2}$ is high enough, what we find implies that we have a good quasiparticle state. We will come later to what $n_{1/2}$ comes out to be. As noted above, this means what we interpret as the quasiparticle mass or the ``Landau mass" (see later)  goes as
\be
m^L\sim f_\pi\sim \la\chi\ra\sim m_0\label{m_0}
,
\ee
where we have applied the chiral symmetry-scale symmetry locking $f_\pi\approx f_\sigma$ discussed in Ref.~\cite{CT} to dense matter.

\section{Fermi-liquid fixed point theory}
Given the $bs$HLS Lagrangian defined in sliding vacua, we wish to connect what is happening at low density to high density relevant to compact stars. For this purpose, we describe how the EFT Lagrangian fares in nuclear matter at a density in the vicinity of $n_0\approx 0.16$fm$^{-3}$.  We will focus on the long-standing problem of the ``quenching" of $g_A$ in nuclear Gamow-Teller transitions at low momentum transfer.
\subsection{The iso-vector anomalous orbital gyromagnetic ratio $\delta g_l$}\label{induced}
To approach the $g_A$ problem, we first recount how the anomalous orbital gyromagnetic ratio $\delta g_l$ in heavy nuclei can be described with Eq.~(\ref{LOSS}) following \cite{FR}.  To proceed,  we take the Lagrangian Eq.~(\ref{LOSS}) with its ``bare" parameters inherited from QCD  at a matching scale $\Lambda_M\approx \Lambda_\chi= 4\pi f_\pi$. Embedded in nuclear medium at density $n$,  the Lagrangian carries the IDDs as discussed above, mainly in terms of the dilaton condensate which slides with density, $f_\sigma^\ast=\la \chi\ra^\ast$~\cite{bshls}. Since in practice the RG decimation is done at a scale lower than $\Lambda_\chi$, there will be ``induced density dependence" termed DD$_{\rm induced}$ that arises when terms that are of higher scale than the effective scale defined for doing RG decimations $\Lambda^\ast <\Lambda_M$, such as short-ranged $n$-body forces for $n>2$ involving vector-meson exchanges, are integrated out. This contribution figures in accounting for the fact that Lorentz invariance is spontaneously broken in nuclear medium. Unless otherwise specified, in what follows, the density dependence containing both IDD and DD$_{\rm induced}$ will be denoted as IDD$^\star$.  In Ref.~\cite{song}, the Lagrangian Eq.~(\ref{LOSS}) endowed with the IDD$^\star$ treated in the mean-field approximation was found to satisfactorily describe the bulk property of nuclear matter, in fact, as well as the relativistic mean-field approaches currently popular in the field. What is notable is that it also gives a fairly accurate description of static quantities.

It turns out that highly relevant to the $g_A$ problem is the anomalous gyromagnetic ratio of the proton in heavy nuclei to which we turn. They are in fact intimately related.

When coupled to the electroweak fields,  the isoscalar gyromagnetic ratio comes out correctly, $g_l^{\rm isoscalar}=1/2$, despite that the nucleon mass  in the Lagrangian scales with density as $m_N^\ast$.  This comes about  by the charge conservation, equivalent to Kohn's theorem for the cyclotron frequency in the electron gas system~\cite{kohn}. What is not trivial at all is that for the iso-vector gyromagnetic ratio, it reproduces~\cite{FR} exactly Migdal's formula~\cite{migdal},
\be
g_l^{\rm isovector}=\Big(\frac 12+ \frac 16\left(\tilde{F}_1^\prime-\tilde{F}_1\right)\Big)\tau_3
,
\ee
with $\tilde{F}=(m_N/m_L)F$, where $F$ is the Landau  parameter and $m_L$ is the Landau quasiparticle mass.  In terms of the Landau fixed point quantities, the anomalous gyromagnetic ratio $\delta g_l$ is then  given by
\be
\delta g_l=\frac 16\big(\tilde{F}_1^\prime-\tilde{F}_1\big)\tau_3.\label{migdal}
\ee
From the Lagrangian Eq.~(\ref{LOSS}) treated at the mean-field level, one can write Eq.~(\ref{migdal}) in terms of the  IDD$^\star$-dependent parameters of Eq.~(\ref{LOSS})~\cite{FR},
\be
\delta g_l=\frac 49\left(\Phi^{-1}-1-\frac 12\tilde{F}_1^\pi\right)\tau_3,
\label{delta-gl}
\ee
where
\be
\Phi=\la\chi\ra^\ast/\la\chi\ra
,
\ee
and $\tilde{F}_1^\pi$ is the pion Fock contribution to the Landau parameter $F_1^\prime$.  It is important to note  that both  $\tilde{F}_1$ and $\Phi$ are RG fixed-point quantities. The former, Fock term, is one-pion-exchange term representing the ``back-flow" current effect that corresponds to $1/N$ correction in the large $N$ expansion where $N = k_F/(\Lambda^\ast -k_F)$  is the cutoff relative to the Fermi surface~\cite{shankar}. It is, however, nonnegligible due to the ``chiral filtering" due to soft pions applicable to isovector M1 operator~\cite{KDR,MR91}.  More on this matter below. The quantity $\Phi$ is related to the Landau-Migdal interaction constant $F_1$ that is controlled  by IDD but can receive corrections from  DD$_{\rm induced}$.
Both quantities in Eq.~(\ref{delta-gl}) are very well known.  At normal nuclear matter density, from deeply bound pionic system~\cite{yamazaki}, $\Phi (n_0)$ is determined as

\be
\Phi (n_0)\simeq 0.80.
\label{eq:Phin0}
\ee
And, $\tilde{F}_1^\pi$ is precisely given by the pionic quantities as
\be
\tilde{F}_1^\pi (n_0)= -0.459.
\label{eq:F1n0}
\ee
This predicted $\delta g_l^p=0.21$ for proton which is in excellent agreement with the experiment value measured in heavy nuclei $\delta g_l^{\rm exp}=0.23\pm 0.03$~\cite{schumacher}.

There are three crucial points to emphasize here which will figure crucially in the quenched $g_A$ problem. One is that the quantity Eq.~(\ref{migdal}) represents the response to the external electromagnetic field with zero momentum transfer of a quasiparticle sitting on the Fermi surface. As such it corresponds to the full quantal response function expressed in terms of {\it simple} shell-model matrix element. Second it involves entirely  RG fixed point quantities, with the assumption that $1/N$ corrections are negligible, which is supported by the result. And third is that the LOSS approximation rids of the dependence on the anomalous dimension $\beta^\prime$ in $\delta g_l$. Therefore, that the $\delta g_l$ is correctly reproduced can be taken as a support for the Fermi-liquid fixed point structure and the LOSS approximation.  We apply this to the $g_A$ problem.

\subsection{Quenching of $g_A$ in nuclei}

Consider Gamow-Teller transitions at zero momentum transfer.  This then deals with a process appropriate for soft-pion theorems. It may not apply, therefore, to Gamow-Teller transitions involving momentum transfer  of order $\sim 100$ MeV in double-$\beta$ decays.

It has been known since 1970's~\cite{wilkinson} that when described in simple shell model, the effective coupling constant for the Gamow-Teller transitions at zero momentum transfer is quenched to a ``universal" value~\footnote{Just to have an idea of what sort of range of $g_A^{\rm eff}$ is involved in light nuclei for which simple shell model calculations have been done with confidence, it is fair to say that  the reasonable value is $g_A^{\rm eff}=1.0 \pm 0.2$.}
 \be
g_A^{\rm eff} \approx 1.
\ee
There have been several papers, both theoretical and experimental,  written on this matter (for up-to-date review, see  Ref.~\cite{gA-review}). Currently it has become particularly prominent for double-$\beta$ decay processes relevant for going beyond the standard model of particle physics. The questions raised were whether the effect is associated with the fundamental QCD issue of how $g_A$ reflects the manifestation of chiral symmetry in nuclear medium and/or the possible role of meson-exchange currents and the intervention of excited baryon degrees of freedom, particularly   $\Delta (1232)$, or  strong nuclear correlations --- involving nucleons only --- requiring going beyond  simple shell-model approaches. Whatever the case may be,  it is fair to say that up to now, what makes $g_A$, which is $g_A=1.2755(11)$ in free space, quenched more or less ``universally" to $\approx 1$ remains still a mystery begging for a simple  answer. Here we describe one way --- perhaps not necessarily the only way --- to understand it. We do this in terms of the LOSS approximation developed above, phrased in Fermi-liquid fixed point theory. An advantage of this way of explaining it is its extreme simplicity while encompassing a wide range of strong nuclear correlation effects and more significantly its possible novel aspect in nuclear interactions.  There are basically no new calculations to be done as all the results are more or less available from the past work.  What's new, however, is that it makes an important conceptual link to what must take place in dense compact star matter that has not yet been  explored up to date.

The relevant part of the Lagrangian Eq.~(\ref{LOSS}) for the problem involving the nucleon field $N$ is
\be
 {\cal L} &=&i\overline{N} \gamma^\mu \del_\mu N -\frac{\chi}{f_\chi}m_N \overline{N}N +g_A \overline{N}\gamma^\mu\gamma_5 \tau_a N{\cal A}_{\mu}^a+\cdots
 ,
 \nonumber\\
 \label{LAG}
 \ee
 where ${\cal A}_\mu$ is the external axial field.
While the kinetic energy term, and particularly the nucleon coupling to the axial field, are scale-invariant by themselves and hence do not couple to the conformal compensator field, the nucleon mass term is multiplied by it. Embedded in the nuclear matter background, the ``bare" parameters of the Lagrangian will pick up the medium VeV. Thus, in Eq.~(\ref{LAG}) the nucleon mass parameter will scale in density, while, most significantly, $g_A$ will remain {\it unscaled},
 \be
 m_N^\ast/m_N &=& \la\chi\ra^\ast/\la\chi\ra\equiv \Phi
 ,
 \label{Phiscaling}\\
 g_A^\ast/g_A &=& 1\label{gA}.
 \ee
The dilaton condensate gets modified when the vacuum is warped by density. It is principally IDD inherited from QCD valid at $n\lsim 2n_0$ with less important contribution from  DD$_{\rm induced}$ as mentioned above~\cite{bshls}. The  relation Eq.~(\ref{gA}) is new and significant  and says that the Lorentz-invariant axial coupling constant {\it does not} scale in density. This result was already indicated in the Skyrme term of the Skyrme model~\cite{BR91} but what is given here is more directly linked to QCD symmetries. Combined with the ``chiral filtering mechanism" to be specified below,  this is the most important  point for the $g_A$ problem.

Now in medium, Lorentz invariance is spontaneously broken, which means that the space component, $g_A^{\rm s}$ and the time component $g_A^{\rm t}$ could be different. Indeed, writing out the space and time components of the nuclear axial current operators, one obtains
 \be
\vec{J}_A^{\pm} (\vec{x}) &=& g^{\rm s}_A \sum_i \tau_i^{\pm} \vec{\sigma}_i \delta(\vec{x}-\vec{x}_i),\label{GT}\\
 J_{5}^{0\pm} (\vec{x})&=&- g^{\rm t}_A \sum_i\tau_i^\pm  \vec{\sigma}_i \cdot (\vec{p}_i  - \vec{k}/2) /m_N \delta(\vec{x}-\vec{x}_i),
 \label{axialcharge}
 \ee
 where $\vec{p}_i$ is the initial momentum of the nucleon making the transition and $\vec{k}$ is the momentum carried by the axial current.  In writing Eqs.~(\ref{GT}) and (\ref{axialcharge}),  the nonrelativistic approximation is made for the nucleon. This approximation is valid not only near $n_0$ but also in the density regime $n\gsim n_{1/2}\approx 2n_0$. This is because the nucleon mass never decreases much after the parity-doubling sets in at $n\approx n_{1/2}$ at which $m_N^\ast\to m_0 \approx (0.6-0.9) m_N$~\cite{bshls}. It will be related to the dilaton condensate  below.

 A simple calculation taking into account Eqs.~(\ref{Phiscaling}) and (\ref{gA}) gives
\be
g_A^{\rm s}=g_A, \ \ g_A^{\rm t}=g_A/\Phi
,
\label{main}
\ee
with $\Phi$ given by Eq.~(\ref{Phiscaling}).

\subsubsection{ Chiral filtering effect}
To confront with nature with high precision, we need two ingredients: (1) accurate nuclear wave functions; (2) reliable nuclear weak currents. In a systematic EFT calculations, the two are to be treated on the same footing in terms of  the scale-chiral power counting.

To proceed, let us suppose that the wave functions are accurately calculable with an accurate potential. It is for the point (2) that the soft-pion theorems figure crucially. In the scale-chiral counting with the scheme espoused in this paper (which is essentially equivalent to chiral counting~\cite{TSP}), with  the axial current taken to be a ``soft pion," there is a soft-pion exchange current that involves double soft-pions coupling to the nucleon dictated by the current algebras. This term was shown to be the most important exchange current contribution to axial charge transitions in nuclei. This was shown first in 1978 using soft-pion theorems~\cite{KDR} and then much later in 1991 using chiral perturbation theory~\cite{MR91}. This  phenomenon was dubbed then ``chiral filter hypothesis" because at the time the high-powered computational techniques currently developed were not available to check quantitatively the arguments involved. The prediction based on the ``hypothesis" was that there would be (a) a huge two-body meson exchange correction due to soft pions to the one-body charge operator Eq.~(\ref{axialcharge}) governing first-forbidden $\beta$ transitions,  with higher chiral-order corrections strongly --- at least by three chiral orders relative to the leading order  --- suppressed, and (b)  the Gamow-Teller transitions, in stark contrast, unprotected by chiral filtering and with the soft pions suppressed, will be, unless accidentally suppressed by symmetries or kinematics,  given {\it predominantly} by the leading order one-body operator Eq.~(\ref{GT}). It was argued~\cite{KDR,MR91} that with no guidance from soft-pion theorems of chiral symmetry, there is no reason to think that exchange-current corrections coming at a few manageable higher chiral orders can be trusted. In other words, chiral symmetry will be of no help tor estimating higher-order  corrections if they are nonnegligible at the given (low) order. As discussed below, there can enter the mixing to the states of higher excitation energies, $\approx 300$ MeV, such as highly correlated nuclear states and $\Delta$-hole states strongly coupled by the nuclear tensor force compounded with many-body current operators.

In short, at low momentum transfers, the Gamow-Teller transitions should be dominantly given by  the single-particle operator with exchange-current contributions strongly suppressed, whereas the axial charge (first-forbidden) transitions should be strongly enhanced by soft-pion processes. This phenomenon is aptly illustrated in the chiral filter phenomenon that involves two sides of the same coin.

Before showing the two sides of the same coin for the axial transitions, it should be mentioned that a soft-pion exchange two-body current is also present for the iso-vector magnetic dipole (M1) transitions in nuclei~\cite{KDR}. In fact it is responsible for the precision calculation of the {\it enhanced} thermal np capture $n+p\to d+\gamma$~\cite{npcapture}.  Since the single-particle operator for the M1 process is of the form that is isospin-rotated from the Gamow-Teller operator, both the Gamow-Teller and M1 operators have in the past been treated in the same way in addressing the issue of quenching.  We should point out that this is not entirely correct.\footnote{ The notable examples are Refs.~\cite{corepolarization} and \cite{bohr-mottelson}.}

Let  us now look at one side of the coin that has to do with the axial charge operator, which is shown to be strongly enhanced by soft-pion effects. The enhancement factor for the axial-charge operator is very simple to calculate. It has been detailed in the published papers but we repeat it succinctly here to stress the robustness of the chiral filter argument.

First involving the soft-pion exchange, the long-ranged exchange axial-charge operator is as precisely known as the one-body operator. Therefore, the ratio of the two-body over one-body matrix elements $R$ can be computed almost nuclear model independently. It comes out to be
\be
R=0.5\pm 0.1\ {\rm for}\   A=12-208
.
\ee
An extremely simple calculation shows that with the two-body effect taken into account,  the effective axial-charge operator is obtained by making the replacement in Eq.~(\ref{axialcharge}) by~\cite{KR}
\be
g_A^t\rightarrow g_A^{t\ast} = \epsilon g_A
,
\ee
with
\be
\epsilon=\Phi^{-1} (1+R/\Phi).
\ee
With $\Phi(n_0)\simeq 0.8$ and $R(n_0)\simeq 0.5$, one predicts
\be
\epsilon (n_0)\approx 2.03.
\ee
This is confirmed unambiguously by the experimental value found in Pb nuclei~\cite{warburton}, $\epsilon^{\rm exp} (n_0)=2.01\pm 0.05$.  The results in $A=12, 16$~\cite{e-MEC} are compatible with the Pb result. This is one of the largest meson-exchange effects known in nuclei and confirms the validity of one side of the coin of the chiral filter hypothesis.

\subsubsection{The other side of the coin}

Now we turn to the other side of the coin of the chiral filter mechanism, i.e., the Gamow-Teller coupling constant. Here the soft-pions are rendered powerless and hence whatever corrections that come to the leading one-body operator must be suppressed  at least by $O(p^2)$ relative to the leading $O(1)$ operator.  This means that higher chiral order corrections in the exchange currents for Gamow-Teller transitions are either strongly suppressed or if not suppressed by kinematics, then cannot be trusted if limited to only a few manageable higher-order terms. In the latter case, to repeat what is said above, one would have to go to much higher order terms than available in the literature, N$^4$LO~\cite{TSP}. Since nuclear EFT can claim accuracy {\it only} for momenta commensurate with soft pions, the higher-order terms, if not suppressed by softness, cannot be calculated with reliable accuracy. This is almost obvious from what other terms can come in the chiral counting as clearly indicated in ~\cite{TSP}. Since one is forced to nonrelativistic approximations, there can then be terms that are not aided by the symmetry, i.e., chiral symmetry, such as for instance recoil terms~\cite{chemtob-rho}, etc., that could make precision calculations of the corrections unfeasible.

What may be involved can be seen in the recent  quantum Monte Carlo calculations in $A=6$ -- $10$ nuclei by Pastore {\it et al}~\cite{monte carlo} done up to N$^4$LO. At that order, the chiral filter is seen to work remarkably, say, at the level of $\lsim 3\%$.  The higher order corrections to the leading sing-particle operator are suppressed, amounting at most to only a few percent. Furthermore there is no indication for $g_A$ quenching, which means that it is the high-order nuclear correlations in the wave functions, not a basic modification of the axial current, that could have been responsible for the ``$g_A$ problem."  If that is the case, then this calculation provides a support for the prediction that $g_A$ should not be affected by the chiral condensate decreasing with density as given by Eq.~(\ref{main}).

There have been papers in which higher order terms in the two-body Gamow-Teller operators are claimed to be substantial at higher density~\cite{jansen} and at higher momentum transfers~\cite{menendez}. It is difficult to assess how model independent these calculations are. There is in fact a counter-example to the higher density effect, found to be substantial at higher density,  of Ref.~\cite{jansen}. In Ref.~\cite{holt}, the Gamow-Teller matrix element in the C14 dating with the two-body corrections to the one-body operator set equal to zero, with however the IDD$^\star$ effects mentioned above taken into account, works extremely  well. There the Gamow-Teller matrix element zeros in on the vicinity of $n_0$. What figures there is the IDD$^\star$ in the tensor force controlling the interaction. This is an  indication that the higher-density effect in the two-body currents that are taken in Ref.~\cite{jansen} is at odds with the result of Ref.~\cite{holt}, which is in agreement with the chiral filter mechanism. As for the case of Ref.~\cite{menendez}, the double-$\beta$ decay involved probes the momentum scale of $\approx 100$ MeV.  This is a hard pion process and hence the chiral filter does not protect it.  Therefore, the exchange current terms to N$^4$LO have no reason to be reliable. In fact, there are cases involving hard pions where low-order chiral perturbation theory fails badly.  The model independence brought in by low-energy theorems of chiral symmetry is lost in that case. As will be discussed below, $\Delta$-hole excitations could also play an important  role there.

\subsubsection{Landau-Migdal Fermi fixed point ${g_A^L}$}
Given the above argument for nonscaling of $g_A$ --- including nondependence on $\beta^\prime$ --- in Gamow-Teller transitions, the question that remains is why  $g_A$ is quenched ``universally" by $\approx 20\%$ in the nuclear shell model calculations.

We offer a simple answer in terms of Landau Fermi-liquid fixed point theory using the scale-chiral EFT Lagrangian, $bs$HLS. The key ingredient for this, we stress, is that the mean-field approximation with the $bs$HLS Lagrangian {\it endowed with the IDDs inherited from QCD}  together with DD$_{\rm induced}$ corresponds to Landau-Fermi liquid fixed point theory \`a la Wilsonian renormalization group to many-fermion systems with Fermi surface ~\cite{bshls}. In the large $N$ limit ($N=k_F/(\Lambda^\ast-k_F)$ where $\Lambda^\ast$ is the cutoff on top of the Fermi surface),  the Landau mass $m_L$ and quasiparticle interactions $\cal {F}$ are at the fixed point, with $1/N$ corrections suppressed~\cite{shankar}. The relation between the Landau mass $m_L$ and the effective $g_A^{\rm L}$, both taken at the fixed point,  is given by~\cite{BR91,FR}
\be
\frac{m_L}{m_N}=1+\frac{1}{F_1}=\left(1-\frac{\tilde{F}_1}{3}\right)^{-1}\approx \Phi \sqrt{\frac{g_A^{\rm L}}{g_A}}.
\ee
Applying the mean-field argument, this relation gives
\be
\frac{g_A^{\rm L}}{g_A}\approx \left(1-\frac 13 \Phi\tilde{F}_1^\pi\right)^{-2}.
\ee
Note that both $\Phi$ and  $\tilde{F}_1^\pi$ are the same RG fixed point quantities that  give the anomalous orbital gyromagnetic ratio $\delta g_l^p$~\cite{FR}.

Let us consider the $g_A^{\rm L}$ at nuclear matter density.  With the values given in Eqs.~\eqref{eq:Phin0} and~\eqref{eq:F1n0} and the current value of $g_A=1.2755$~\cite{gA} we get 
\be
g_A^{\rm L} (n_0) \approx 0.79 g_A\approx 1.0.\label{landaugA}
\ee
This is precisely $g_A^{\rm eff}$ needed in the shell-model calculations~\cite{gA-review} and in the giant Gamow-Teller resonances~\cite{sakai}.  Note here the crucial role of the pionic contribution  interlocked with the dilaton condensate for the quenching. It turns out that the density dependence in $\Phi$ (dropping with density) nearly cancels the density dependence in
$\tilde{F}_1^{\pi}$ (increasing with density) so that the product  $\Phi\tilde{F}_1^\pi$ becomes nearly independent of density: The $g^L_A$ differs by less than 2\% between the densities $\frac 12 n_0$ and $n_0$.  Thus, what we could call  ``Landau-Migdal $g_A$" Eq.~(\ref{landaugA}), although evaluated for nuclear matter,  is highly robust and nuclear-model independent.  Hence, it should apply to nuclear matter as well as to finite nuclei, both light and heavy.

Now how does this $g_A^{\rm L}$ correspond to $g_A^{\rm eff}$ in the shell-model calculations?

To answer this question, recall that {\it at the Fermi-liquid fixed point in our formulation with IDD$^\ast$ suitably implemented,  the $\beta$ functions for the quasiparticle interactions $\cal{F}$, the mass $m_L$, the Gamow-Teller coupling $g_A^{\rm L}$, etc., at a given density  should be suppressed by $1/N$}. This means in particular that the {\it quasiparticle} loop corrections to the effective $g_A$ should be suppressed. It is therefore the effective coupling constant, duly implemented with density-dependent condensates inherited from QCD and with high-order quasiparticle correlations subsumed, to be applied to noninteracting quasiparticles, that is,  simple particle-hole configurations --- without correlations --- in shell model calculations.  This corresponds {\it effectively} to what is captured {\it microscopically} in the {\it ab initio} quantum Monte Carlo calculation of Ref.~\cite{monte carlo}. This is the analog to the gyromagnetic ratio for the quasiparticle sitting on top of the Fermi sea. That it agrees precisely with the experiment implies that the approximation is good.

Now suppose that no LOSS approximation were made to the CT Lagrangian at the leading chiral order. Then $g_A^{\rm eff}$ would have the $\beta^\prime$ dependence of the form
\be
g_A (\beta^\prime,c_A) =\big(c_A+(1-c_A)\Phi^{\beta^\prime}\big) g_A^{\rm L}.
\ee
Suppose that $c_A\neq 1$ and $\beta^\prime\neq 0$. This can involve an intricate interplay of the two parameters. It would be interesting to explore whether precision calculations of the Gamow-Teller transitions can give information in medium on $c_A$ and  $\beta^\prime$, the basic property of scale-chiral symmetry in baryonic matter.

In what follows we will continue with the LOSS approximation.
\subsubsection{$\Delta$-hole contributions to $g_A$}
In identifying the Landau-Migdal  fixed-point quantity $g_A^L$ with $g_A^{\rm eff}$, we have assumed that the mean-field approximation with Eq.~(\ref{LOSS}) with the sliding vacua encoded therein is equivalent to Fermi-liquid fixed point theory built on top of the Fermi sea characterized by the Fermi momentum $k_F$ or density $n$.  There the Landau interaction parameters are given in terms of four-Fermi interactions involving nucleon quasiparticles. The underlying premise is that nucleons are the appropriate baryon degrees of freedom and $n$-body interactions for $n>2$ are suppressed by $1/N$. The Gamow-Teller states are strongly excited from the ground states by the nuclear tensor forces, the virtual excited states being with excitation energy $\approx (200-300)$ MeV. $\Delta$-hole states at excitation energy $\approx 300$ MeV can equally be excited by the tensor force involving $\Delta$N channels. If one were to do a microscopic calculation, then one cannot ignore  $\Delta$-hole states excited by the $\Delta$N tensor force, given that one must include nucleon particle-hole states excited by the NN tensor force for core polarizations involving nucleon degrees of freedom of comparable excitation energies.

That $\Delta$-hole state could screen $g_A$ has been understood since early 1970's~\cite{delta-hole} and invoked for giant Gamow-Teller resonances~\cite{Delta-GTR,bohr-mottelson}. If one extends the configuration space to the $\Delta$ sector, then the $\Delta$-hole contributions can be considered as a part of generalized core polarization produced strongly by tensor forces. One could also phrase this in terms of a generalized Landau(-Migdal) Fermi-liquid theory involving $g_0^\prime$ interaction parameters in NN and N$\Delta$ channels. See Ref.~\cite{ichimura} for review.

It should be emphasized that seen from the point of view of $bs$HLS, what has been done up to date in microscopic calculations is  incomplete due to the fact that the crucial interplay of the $\pi$ and $\rho$ tensors --- e.g., suppressed tensor force at higher density  as seen in Fig.~\ref{tensor} --- has been missing in the works done so far. As far as we know, the close interplay in the tensor force between the pionic and $\rho$ tensors has not been taken into account in {\it ab initio} calculations used in the potentials based on chiral Lagrangians. The sophisticated phenomenological potentials such as the one used in Ref.~\cite{monte carlo} do not contain this crucial feature. The same applies to the $\Delta$-hole ``core polarization" calculation \cite{delta-hole} as indicated in the weaker Landau-Migdal  parameter $g_0^\prime$ required in the $\Delta$-N channel~\cite{ichimura}.   A full microscopic calculation of the type performed in \cite{monte carlo} --- which presently gives support to the chiral filter hypothesis and the unimportant role of $\Delta$-hole states  --- that duly takes into account the cancellation effect in the tensor forces, both in the NN and $\Delta$N channels,  would ne needed to settle this matter.

We should reiterate the underlying theme that in the mean-field treatment of $bs$HLS in the LOSS approximation that {\it albeit approximately} corresponds to the single-decimation Fermi-liquid fixed-point theory,  what we call Landau-Midal $g_A^L$ is to encode the full nuclear correlations for the quasiparticle sitting on top of the Fermi sea and hence is to represent the simple shell model matrix element with the effective $g_A$.  What is involved is the IDD$^\ast$s defined at a given cutoff for decimation and the fixed-point properties in the parameters that figure in it, i.e., $\Phi$, $m_L$ and $F_1^\pi$.  Since the baryon resonance degrees of freedom are supposed to have been integrated out in defining $bs$HLS where $\Delta$ is absent,  their effects should in principle be encoded in the ``bare" parameters with IDD$^\ast$s.  The result Eq.~(\ref{landaugA}) therefore is the whole story. There is no fundamental quenching, and most likely no significant quenching coming from meson-exchange currents.

\section{Dense Baryonic Matter}
So far we have established that up to nuclear matter density, long wave-length fluctuations on top of the Fermi surface can be well described with the LOSS Lagrangian Eq.~(\ref{LOSS}). The bulk property of normal nuclear matter has  also been fairly well tested~\cite{song}. Let us now go to higher density, say, density greater than that of topology change $n_{1/2}$. The skyrmion matter description with scale-invariant HLS theory, $s$HLS, does not allow to pin down the location of $n_{1/2}$. Although the skyrmion-half-skyrmion transition is robust, dependent only on topology associated  with the pion, the location of $n_{1/2}$ depends significantly on what other (matter) degrees of freedom are included. Clearly  $n_{1/2}$ cannot be lower than  the normal nuclear matter at $n_0$,  but it cannot be much higher either since then quark-gluon degree of freedom involving quark percolation must intervene explicitly. Analyses made in Ref.~\cite{bshls} indicate for consistency with Nature that the reasonable density is $n_{1/2}\approx 2n_0$. This is also reasonable in considering the possible overlap of the topology change with the presence of quark-gluon degrees of freedom in the density range $\approx (2-4)n_0$ as proposed in Ref.~\cite{baymetal}.\footnote{It has been suggested  that quarks/gluons could be traded in at density $\approx 2n_0$  for topology along the line of reasoning based on the Cheshire Cat principle~\cite{cndIII}.}
At present there is no known way, experimental or theoretical, to pin down the value for $n_{1/2}$. But we will suggest below and in a forthcoming paper that the tidal deformability to be measured in gravitational waves coming from coalescing neutron stars could provide the means to do so.

In what follows we will simply take  $n_{1/2}=2n_0$ as was done in Ref.~\cite{bshls}.

We approach the problem in two ways. First the single-decimation RG approach which was used above for long wave-length fluctuations and the $V_{\rm lowk}$ RG approach that corresponds to the double decimation RG approach, which takes into account certain $1/N$ corrections. The latter has been tested in various ways as described in Ref.~\cite{bshls} and the references given therein. We will not repeat any details here. For the EoS we are interested in, the $V_{\rm lowk}$RG is indispensable.

While the single-decimation approach resorted to the mean field treatment, the $V_{\rm lowk}$ approach employs the nuclear potentials built with the $bs$HLS Lagrangian. The potentials are constructed on the sliding vacua and therefore have the IDD$^\star$s encoded in the Lagrangian. Our basic premise, stated previously, is that doing the double-decimation $V_{\rm lowk}$RG is equivalent to doing {\it full} Landau Fermi-liquid theory going beyond the fixed-point approximation~\cite{SB-fermiliquid}.  For both approaches, the topology change discussed in Section \ref{topology} defines the IDD$^\star$s at the density $n_{1/2}$. The ``bare" parameters in $bs$HLS Eq.~(\ref{LOSS}) satisfy what could be called ``master formula" of the form~\cite{bshls}
\begin{equation}
\frac{m_N^\ast}{m_N} \approx \frac{g_{\rho,\omega}}{g_{\rho,\omega}^\ast} \frac{m_{\rho,\omega}^\ast}{m_{\rho,\omega}} \approx \frac{m_\sigma^\ast}{m_\sigma} \approx \left( \frac{m_\pi^\ast}{m_\pi} \right)^2 \approx \frac{f_\pi^\ast}{f_\pi} \approx \frac{\langle \chi \rangle^\ast}{\la\chi\ra}\equiv \Phi\,,\label{master}
\end{equation}
where
\begin{equation}
m_\sigma^{\ast\,2} \equiv  \left. \frac{\partial^2 V(\sigma)}{\partial\,\sigma^2} \right|_{\sigma = \langle \sigma \rangle^\ast}\,
,
\end{equation}
with $V$ being  the dilaton potential. The only density-scaling functions that we need to fix are $\Phi_V\equiv g_V^\ast/g_V$ for $V=(\rho,\omega)$ and $\Phi$, which is known for $n\lsim n_{1/2}$ extrapolated from $n_0$ and governed for $n\gsim n_{1/2}$ by the vector manifestation for $\rho$ and unknown for $\omega$, so is in a way a free parameter. With these functions suitably determined, this formula is applied in Ref.~\cite{bshls} to the EoS for both normal nuclear matter and highly dense matter with only small fine tuning. We will not go into details of the calculation but use the results in the discussion that follow. As stated, the EoS so constructed works fairly well and agrees with what is given empirically at least up to $\approx n_0$ and we assume that it is applicable to $\approx n_{1/2}$. It is the density regime $n\gsim n_{1/2}$ to which we turn our attention.

\subsection{Trace anomaly}

That scale symmetry in the chiral limit is broken quantum mechanically in QCD is given by the nonvanishing trace of the energy momentum tensor $\theta_\mu^\nu$. We are interested in knowing how the trace of the energy momentum tensor $\la\theta_\mu^\mu\ra$ behaves in our effective Lagrangian $bs$HLS Eq.~(\ref{LOSS}) in the phase with $n>n_{1/2}$. We look at this first in the mean-field approximation (single-decimation Fermi liquid) and then at the $V_{\rm lowk}$RG (double-decimation Fermi liquid).

\subsubsection{Mean-field approximation}

Expanding out the Lagrangian Eq.~(\ref{LOSS}) and noting that the mean fields of $\pi$ and $\rho$ do not contribute, it is straightforward to calculate the energy density $\epsilon$ and the pressure $P$ and find in the chiral limit
\be
\langle \theta^\mu_\mu \rangle = \epsilon - 3 P
 = 4V(\langle \chi \rangle) - \langle \chi \rangle \left. \frac{\partial V( \chi)}{\partial \chi} \right|_{\chi = \langle \chi \rangle}. \label{TEMT1}
\ee
$\langle \theta^\mu_\mu \rangle$ depends on $\beta^\prime$ but it is also a constant independent of density. Therefore, we arrive at the conclusion that $\langle \theta^\mu_\mu \rangle$ is independent of density. This corresponds to calculating in the Fermi-liquid approach with $1/N$ corrections ignored.

\subsubsection{$V_{\rm lowk}$ RG}

\begin{figure}[h]
\begin{center}
\includegraphics[width=8cm]{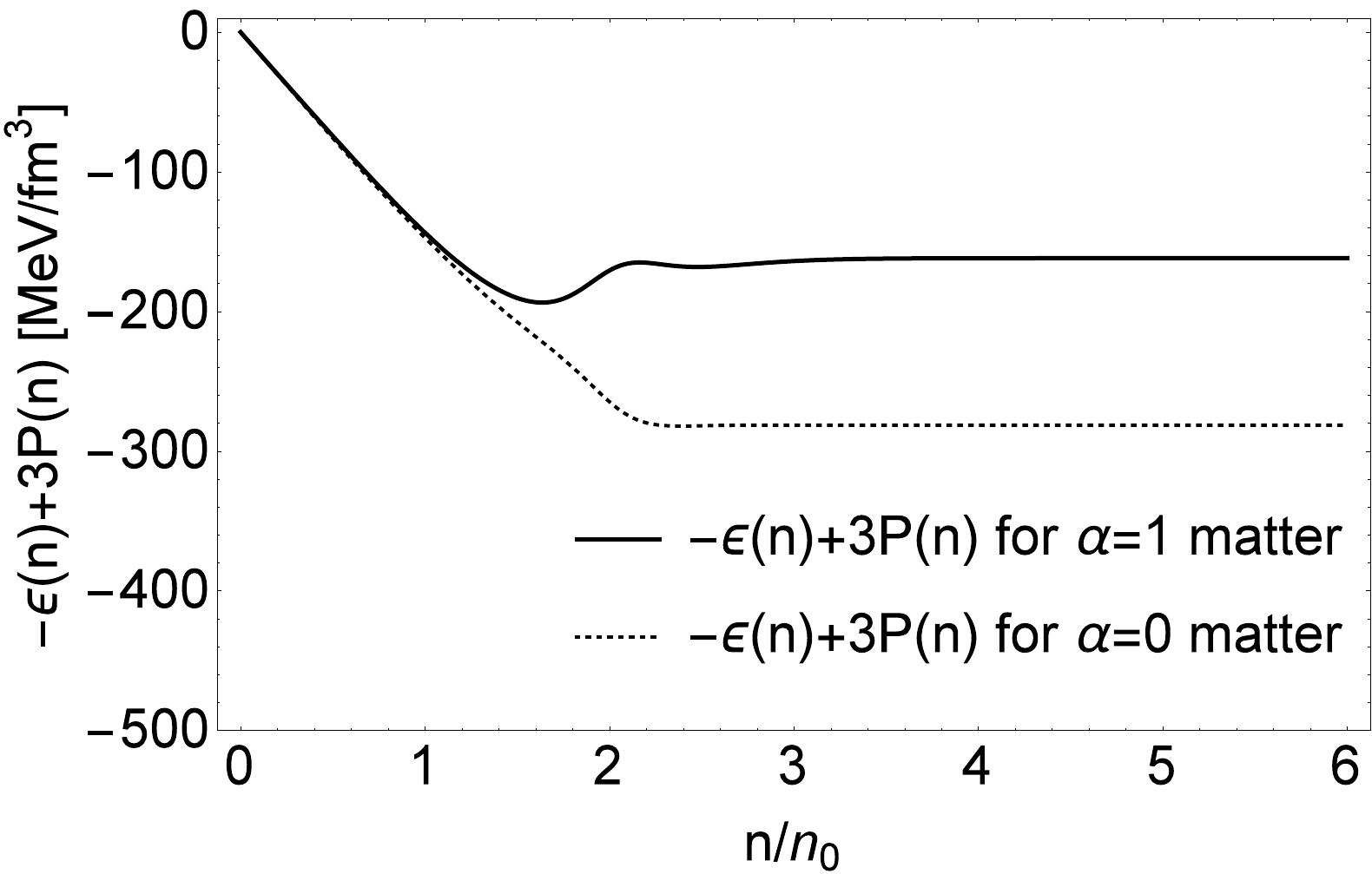}
\caption{ $\la\theta_\mu^\mu\ra$  in $V_{\rm lowk}$ RG reproduced from Ref.~\cite{bshls}.
 }\label{TEMT}
 \end{center}
\end{figure}

Computing $\epsilon$ and $P$ in $V_{\rm lowk}$ that include certain loop corrections going beyond the large $N$ limit, one gets the result given in Fig.~\ref{TEMT}. $\theta_\mu^\mu$ is constant of density for both nuclear and neutron matter, hence in compact-star matter in $\beta$ stability with $\alpha\neq 1$.
It is remarkable that the condensate $\la\chi\ra$ going to a density-independent constant for $n>n_{1/2}$ together with parity-doubling results from scale symmetry encoded in the strong correlations involved.

\subsection{Sound velocity}

If scale symmetry is unbroken, then $\la\theta_\mu^\mu\ra=0$ and hence the sound velocity is $v_s^2=1/3$. This is referred to as ``conformal sound velocity."  Since in the medium-free vacuum, $\la\theta_\mu^\mu\ra\neq 0$ due to the trace anomaly, the conformal velocity is {\it not} present in QCD proper.

\begin{figure}[h]
\begin{center}
\includegraphics[width=8cm]{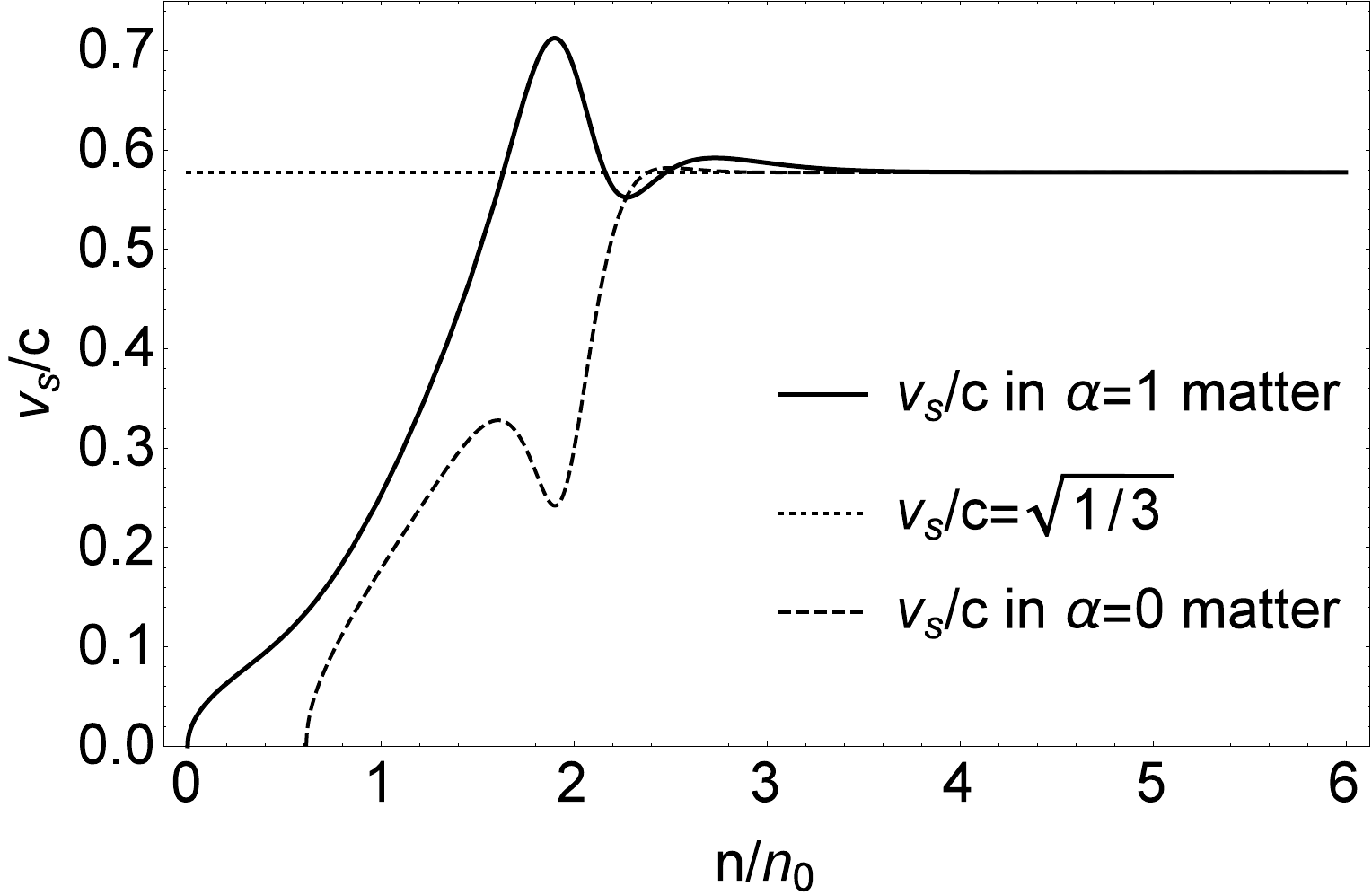}
\caption{ Sound velocity calculated in $V_{\rm lowk}$ RG.
 }\label{sound}
 \end{center}
\end{figure}

Now consider $\la\theta_\mu^\mu\ra$ and the sound velocity in dense medium. The two are related by
\be
\frac{\partial}{\partial n} \la\theta_\mu^\mu\ra=\frac{\partial \epsilon (n)}{\partial n} (1-3{v_s^2})
,
\label{derivTEMT}
\ee
with $v_s^2=\frac{\partial P(n)}{\partial n}/\frac{\partial \epsilon(n)}{\partial n}$. Since  as seen above we have $\la\theta_\mu^\mu\ra$ independent of density for $n\gsim n_{1'2}$ (in both mean field and $V_{\rm lowk}$), the left-hand side of Eq.~(\ref{derivTEMT}) is  zero in that density regime.  If we assume that there is no extremum in the energy density for density $n > n_{1/2}$ --- which we believe is justified,  then $\frac{\partial \epsilon (n)}{\partial n} \neq 0$. It then follows that
\be
v_s^2 = 1/3.
\ee
One can see in Fig.~\ref{sound}  that this indeed is verified in the {numerical} $V_{\rm lowk}$RG calculations~\cite{bshls}.

It should be stressed that  this is {\it not} the genuine conformal velocity, absent in QCD proper. The trace anomaly is still nonzero there. Yet the velocity approaches conformal one as the system goes into a ``phase" where the parity-doubling symmetry emerges. Therefore, we identify it as a ``pseudoconformal sound velocity" associated with a possibly emergent scale symmetry. The viability of this emergent conformal sound velocity was verified in Ref.~\cite{PCmodel} in terms of a ``pseudoconformal EoS."

\subsubsection{Dilaton-limit fixed point}

In Sec.~\ref{parity-doubling}, we noted that in the topological description for the skyrmion-to-half-skyrmion transition, the effective baryonic mass went to a constant at $n_{1/2}$ because of parity doubling symmetry emergent in the transition.  It had to do with the chiral condensate, averaged over space,  going to zero in the half-skyrmion phase, while chiral symmetry is still broken with the pion field excited.  As QCD does not possess this symmetry {\it explicitly}, neither does the effective Lagrangian we constructed.\footnote{Whereas  it does emerge in skyrmion matter at $n\gsim n_{1/2}$.}  To study parity doubling in the mean field, therefore, we need to introduce it in consistency with the symmetries of QCD. This is very much like hidden local symmetry implemented by exploiting the gauge equivalence to nonlinear $\sigma$ model. This can be done along the same line of reasoning done by Ref.~\cite{detar-kunihiro}. In Ref.~\cite{interplay} where previous references are given,  the parity doubling was incorporated into the $bs$HLS, Eq.~(\ref{LOSS}). The baryonic Lagrangian is of the form
\begin{eqnarray}
\mathcal{L}^{N}
&=& \bar{N} i \sbar{D} N - \bar{N}  \hat{\mathcal{M}} N
{}+ g_A\bar{N}\gamma^\mu\hat{G}\hat{\alpha}_{\perp\mu}\gamma_5 N
\nonumber \\
&&
{}+ g_{V\rho}\bar{N}\gamma^{\mu} \hat{\alpha}_{\parallel\mu} N
{}+ g_{V0} \bar{N}\gamma^{\mu}\mbox{Tr}\left[\hat{\alpha}_{\parallel \mu} \right]N\,,
\label{Nlagrangian}
\ee
with
\be
\hat{\mathcal{M}}
&=& \left( \begin{array}{cc} m_{N_+} & 0 \\
0 & m_{N_-} \end{array} \right)\,,
\
\hat{G} = \left( \begin{array}{cc}
\tanh\delta & \gamma_5/\cosh\delta \\
\gamma_5/\cosh\delta & -\tanh\delta
\end{array}\right)\,\nonumber
\end{eqnarray}
and
\begin{eqnarray}
&& m_{N_{\pm}}
= \mp  g_{2} f_{\pi} + \sqrt{\left( g_{1}f_{\pi}\right)^{2} + m_{0}^{2}}\,,
\label{mass}
\\
&& \cosh \delta = \frac{m_{N_+} + m_{N_-} }{2m_{0}}\,.
\end{eqnarray}
Here $N^T=(N_+,N_-)$ is the parity doublet nucleon field, $g_A$, $g_{V\rho}$, $g_{V0}\equiv \frac 12 (g_{V\omega}-g_{V\rho})$ and $g_{1,2}$  are dimensionless parameters and $m_0$ is the chiral invariant baryon mass that remains nonzero in the chiral symmetric phase. In dense medium, the $U(2)$ symmetry for $\rho$ and $\omega$ is broken, so the relevant symmetry is $SU(2)\times U(1)$.

We now wish to look at Eq.~(\ref{LOSS}) with the baryonic Lagrangian replaced by Eq.~(\ref{Nlagrangian}) in what is called ``dilaton limit"~\cite{bira,PLRS}. The idea is essentially the same as going from the nonlinear $\sigma$ model Lagrangian to a scale invariant chiral Lagrangian by dialing a constant of linear $\sigma$ model as discussed by Yamawaki~\cite{yamawaki}. For this purpose,  we do the field reparametrization,
\be
\Sigma=U\chi\frac{f_\pi}{f_\sigma}.
\ee
One finds that if one dials ${\rm Tr} (\Sigma\Sigma^\dagger)\to 0$ in the re-parameterized Lagrangian, there results  a singular part
\begin{eqnarray}
\mathcal{L}_{\rm sing} & =&
\left( 1 -g_A \right) {\cal A} \left( 1/\mbox{Tr} \left[ \Sigma \Sigma^{\dagger} \right]\right) \nonumber\\
&&{} + \left( \delta -1\right) {\cal B} \left( 1/\mbox{Tr} \left[ \Sigma \Sigma^{\dagger} \right]\right)\,,
\label{sing}
\end{eqnarray}
where $\delta \equiv \frac{f_\pi^2}{f_\sigma^2}$ and $({\mathcal A}, {\mathcal B})$ are nonsingular quantities containing $\Sigma$ and baryon field. That ${\mathcal L}_{\rm sing}$ be absent leads to the conditions that
\be
g_A\rightarrow 1,
\quad
f_\pi \to f_\sigma
\ee
This is the ``dilaton limit." Taken in the mean field, the quantities involved will be in-medium quantities and hence should be affixed with $\ast$.

There are two points worth stressing here.

One is that there is no obvious correlation between  $g_A\to 1$ here as density passes $n_{1/2}\approx 2n_0$  going toward the dilaton-limit fixed point and the ``quenching" of $g_A=1.2755$ to $g_A^L\approx 1$ in finite nuclei and nuclear matter described as Fermi-liquid fixed point matter.  It is an intriguing question what happens to $g_A^L\approx 1$ as density increases to the dilaton limit fixed point where $g_A\to 1$. Second, the dilaton-limit fixed point corresponds to $\Sigma\to 0$.  It is not clear whether this limit corresponds to the IR fixed point in the CT theory, which in our LOSS approximation corresponds to $\beta^\prime\to 0$.

\section{Symmetry energy and tidal deformability}

Here we return to the symmetry energy problem discussed in Sec.~\ref{topology}. There how the cusp predicted by the skyrmion crystal description can be qualitatively reproduced by the behavior of the nuclear tensor force caused by the appearance of half-skyrmion phase at $n\approx 2n_0$ and its impact on the vector manifestation property of the $\rho$ meson predicted by the RG analysis with HLS. Here we revisit the same problem with the $V_{\rm lowk}$RG and discuss how the cusp can influence the tidal deformability $\Lambda$.

In the $V_{\rm lowk}$ RG approach~\cite{bshls}\footnote{This approach, with only a few parameters, reproduces all the bulk properties of normal nuclear matter and also the EoS for massive compact stars. For completeness, let us mention a few of them here:  $E_0(n_0)/A-m_N\simeq -15.5 $ MeV, $n_0\simeq 0.154 $ fm$^{-3}$, $K(n_0)\simeq 215 $ MeV, $J=E_{\rm sym}(n_0)\simeq  26$ MeV, $L(n_0)\simeq  49 $ MeV. They are more or less consistent --- although not fine-tuned --- with the presently quoted empirical values $E_0/A-m_N= -15.9 \pm 0.4$ MeV, $n_0= 0.164 \pm 0.007$ fm$^{-3}$, $K= 240\pm 20$ MeV or  $230\pm 40$ MeV, $30\lsim E_{\rm sym} \lsim 35  $ MeV, $20 \lsim L\lsim 66$ MeV. The properties of the massive star are: $M^{\rm max}=2.05 M_\odot$, $R=12.19$ km, $n_{\rm cent}=5.1n_0$. }, the cusp of the symmetry energy is smoothed to a continuous changeover from soft to hard as was shown in Fig.~\ref{Esym}.

The characteristic feature of the cusp, namely the drop of the symmetry energy going toward the changeover density $n_{1/2}$ followed by increase after, could have an impact on properties of processes that occur in the vicinity of the density involved. In fact, the recent gravitational wave observation GW170817~\cite{GW-Love} offers an interesting possibility. The bound for the dimensionless tidal deformability $\Lambda$ reported from the observation in Ref.~\cite{GW-Love} involves $\approx 1.4M_\odot$ star, which in the $V_{lowk}$ RG analysis~\cite{bshls}, supports a central density $\approx 2n_0$, just about the same as as the symmetry energy cusp density $n_{1/2}$. It was found in \cite{bshls} that with $n_{1/2}=2.0n_0$  $\Lambda$ came out $\Lambda\approx 790$, slightly below the bound given by GW170817 $\Lambda < 800$. This would lead to the constraint $n_{1/2}> 2.0n_0$. Given that the most important quantity for the quantity concerned in the EoS is the symmetry energy, a possibility to lower  the theoretical value of $\Lambda$ is to increase the topology change density to above $2 n_0$ which according to Fig.~\ref{tensor} would make $E_{\rm sym}$ more attractive. Other properties of stars, it turns out,  are not sensitive to the exact location of $n_{1/2}$. In fact, it has been verified that very little is changed with $n_{1/2}$ between 1.5$n_0$ and 2.0$n_0$~\cite{dongetal}. So it is very possible that $\Lambda$ could be lowered by exploiting the attraction gained by pushing $n_{1/2}$ somewhat higher without upsetting other quantities.

As noted above, there has been, up to date, no known way, theoretical or experimental, to pin down the topology change density. Fully satisfactory quantized approach to skyrmion matter could do it but it is an unsolved  daunting problem.  It seems possible to pin it down once the precise value for $\Lambda$ is determined in the future from gravitational wave data.
\section{Remarks}

In this paper, we have shown that the long-standing problem of ``quenched" $g_A$ in nuclear shell model can be understood in terms of a quasiparticle on the Fermi surface with a Fermi-liquid fixed point constant $g^L_A$  in long-wavelength response to the external field. There is, in true sense,  {\it no} quenching to  the fundamental constant $g_A$, unlike the pion decay constant, due to the vacuum change in nuclear medium. Many authors~\cite{corepolarization} have argued  since a long time the observed ``quenched $g_A$" simply reflects core-polarizations.   This issue was more recently raised in connection with the role of  $\Delta$-hole configurations in a generalized core-polarization mechanism~\cite{auerbach}. But contrary to what has been generally accepted, the argument does not hold equally for iso-vector M1 matrix elements as pointed out above in connection with the chiral-filter mechanism mentioned above.

What seems to be crucially at work in this phenomenon is the emergence in nuclear matter at low as well as high densities of  scale symmetry hidden in QCD~\cite{CCT}. As density goes above the topology change $n_{1/2}$, the Landau-Migdal constant $g_A^L\approx 1$ goes over to the ``dilaton limit" value $g_A^{\rm dilaton}=1$ at high density. Whether there is any connection and if there is, how this transition takes place is yet to be understood, but what is happening  is that as density goes above $n_{1/2}$,  parity doubling occurs and the sound velocity of the system goes to the conformal value $v_s^2=1/3$ although the trace of the energy momentum tensor is not zero and hence scale symmetry is still broken.  This takes place at a density $\approx 2n_0$, far below asymptotic density, so it is a precocious happening.

It has been argued that the conformal sound velocity cannot be supported unless there is a change of degrees of freedom~\cite{tews}. Our answer to that no-go argument is that there is the topology change from skyrmions to half-skyrmions which essentially changes the relevant degrees of freedom. What we are suggesting is that  the  strong-coupling quark-gluon degrees of freedom hidden in QCD, such as in the scenario proposed by Baym {\it et al}.~\cite{baymetal} at density $\sim (2-4)n_0$, are traded in for topology at the same density regime. Such trade-in mechanism has been seen in various processes  in terms of what is known as ``Cheshire-Cat phenomenon"~\cite{cheshire,cndIII}.

Finally, what is striking  in what we have gotten is that although there is no rigorous proof for the existence of an IR fixed point in QCD for nuclear physics, hidden scale symmetry emerges in the half-skyrmion ``phase" in which parity doubling occurs, which is present buried in finite nuclei and ``unburied" in dense matter.

\subsection*{Acknowledgments}
Y.~L. Ma is supported in part by National Science
Foundation of China (NSFC) under Grants No. 11475071 and No. 11547308 and the Seeds Funding of Jilin
University.


\end{document}